\documentclass{llncs}

\usepackage{graphicx}
\begin{document}

\title{Oscillating Statistical Moments for Speech Polarity Detection}

\author{Thomas Drugman, Thierry Dutoit}

\institute{TCTS Lab, University of Mons, Belgium}
\maketitle

\begin{abstract}
An inversion of the speech polarity may have a dramatic detrimental effect on the performance of various techniques of speech processing. An automatic method for determining the speech polarity (which is dependent upon the recording setup) is thus required as a preliminary step for ensuring the well-behaviour of such techniques. This paper proposes a new approach of polarity detection relying on oscillating statistical moments. These moments have the property to oscillate at the local fundamental frequency and to exhibit a phase shift which depends on the speech polarity. This dependency stems from the introduction of non-linearity or higher-order statistics in the moment calculation. The resulting method is shown on 10 speech corpora to provide a substantial improvement compared to state-of-the-art techniques.

\end{abstract}

\section{Introduction}
The polarity of speech may affect the performance of several speech processing applications. This polarity arises from the asymmetric glottal waveform exciting the vocal tract resonances. Indeed, the source excitation signal produced by the vocal folds generally presents, during the production of voiced sounds, a clear discontinuity occuring at the Glottal Closure Instant (GCI, \cite{Drugman-GCI}). This discontinuity is reflected in the glottal flow derivative by a peak delimitating the boundary between the glottal open phase and return phase. Polarity is said positive if this peak at the GCI is negative, like in the usual representation of the glottal flow derivative, such as in the Liljencrant-Fant (LF) model \cite{LF}. In the opposite case, polarity is negative.

When speech is recorded by a microphone, an inversion of the electrical connections causes the inversion of the speech polarity. Human ear is known to be insensitive to such a polarity change \cite{Sakaguchi}. However, this may have a dramatic detrimental effect on the performance of various techniques of speech processing. In unit selection based speech synthesis \cite{NUU}, speech is generated by the concatenation of segments selected from a large corpus. This corpus may have been built through various sessions, possibly using different devices, and may therefore be made of speech segments with different polarities. The concatenation of two speech units with different polarity results in a phase discontinuity, which may significantly degrade the perceptual quality when taking place in voiced segments of sufficient energy \cite{Sakaguchi}. There are also several synthesis techniques using a pitch-synchronous overlap-add (PSOLA) which suffer from the same polarity sensitivity. This is the case of the well-know Time-Domain PSOLA (TDPSOLA, \cite{TDPSOLA}) method for pitch modification purpose.

Besides, efficient techniques of glottal analysis require to process pitch synchronous speech frames. For example, the three best approaches considered in \cite{Drugman-GCI} for the automatic detection of GCI locations, are dependent upon the speech polarity. An error on its determination results in a severe impact on the reliability and accuracy performance. There are also some methods of glottal flow estimation and for its parameterization in the time domain which assume a positive speech polarity \cite{Drugman-GF}.

This paper proposes a new approach for the automatic detection of speech polarity which is based on the phase shift between two oscillating signals derived from the speech waveform. Two ways are suggested to obtain these two oscillating statistical moments. One uses non-linearity, and the other exploits higher-order statistics. In both cases, one oscillating signal is computed with an \emph{odd} non-linearity or statistics order (and is \emph{dependent} on the polarity), while the second oscillating signal is calculated for an \emph{even} non-linearity or statistics order (and is \emph{independent} on the polarity). These two signals are shown to evolve at the local fundamental frequency and have consequently a phase shift which depends on the speech polarity.

This paper is structured as follows. Section \ref{sec:ExMethods} gives a brief review on the existing techniques for speech polarity detection. The proposed approach is detailed in Section \ref{sec:Approach}. A comprehensive evaluation of these methods is given in Section \ref{sec:Experiments}, providing an objective comparison on several large databases. Finally Section \ref{sec:conclu} concludes the paper.

\section{Existing Methods}\label{sec:ExMethods}
Very few studies have addressed the problem of speech polarity detection. We here briefly present three state-of-the-art techniques achieving this purpose.

\subsection{Gradient of the Spurious Glottal Waveforms (GSGW)}
The GSGW method \cite{GSGW} focuses on the analysis of the glottal waveform estimated via a framework derived from the Iterative Adaptive Inverse Filtering (IAIF, \cite{IAIF}) technique. This latter signal should present a discontinuity at the GCI whose sign depends on the speech polarity. GSGW therefore uses a criterion based on a sharp gradient of the spurious glottal waveform near the GCI \cite{GSGW}. Relying on this criterion, a decision is taken for each glottal cycle and the final polarity for the speech file is taken via majority decision.

\subsection{Phase Cut (PC)}
The idea of the PC technique \cite{PC} is to search for the position where the two first harmonics are in phase. Since the slopes are related by a factor 2, the intersected phase value $\phi_{cut}$ is:

\begin{equation}
\phi_{cut}=2\cdot \phi_1 - \phi_2,
\end{equation}

where $\phi_1$ and $\phi_2$ denote the phase for the first and second harmonics at the considered analysis time. Assuming a minimal effect of the vocal tract on the phase response at such frequencies, $\phi_{cut}$ closer to 0 (respectively $\pi$) implies a positive (respectively negative) peak in the excitation \cite{PC}. PC then takes a single decision via a majority strategy over all its voiced frames.

\subsection{Relative Phase Shift (RPS)}

The RPS approach \cite{PC} takes advantage from the fact that, for positive peaks in the excitation, phase increments between harmonics are approximately due to the vocal tract contribution. The technique makes use of Relative Phase Shifts (RPS's), denoted $\theta(k)$ and defined as:

\begin{equation}
\theta(k) = \phi_k - k\cdot \phi_1,
\end{equation}

where $\phi_k$ is the instantaneous phase of the $k^{th}$ harmonic. For a positive peak in the excitation, the evolution of RPS's over the frequency is smooth. Such a smooth structure is shown to be sensitive to a polarity inversion \cite{PC}. For this, RPS considers harmonics up to 3kHz, and the final polarity corresponds to the most represented decisions among all voiced frames.

\section{Oscillating Moments-based Polarity Detection (OMPD)}\label{sec:Approach}

In \cite{Drugman-GCI}, we proposed a method of Glottal Closure Instant (GCI) determination which relied on a mean-based signal. This latter signal had the property to oscillate at the local fundamental frequency and allowed good performance in terms of reliability (i.e leading to few misses or false alarms). The key idea of the proposed approach for polarity detection is to use two of such oscillating signals whose phase shift is dependent on the speech polarity. For this, we define the oscillating moment $y_{p_1,p_2}(t)$, depending upon $p_1$ and $p_2$ which respectively are the statistical and non-linearity orders, as:

\begin{equation}\label{eq:OMDP1}
y_{p_1,p_2}(t) = \mu_{p_1}(x_{p_2,t})
\end{equation}

where $\mu_{p_1}(X)$ is the $p_1^{th}$ statistical moment of the random variable $X$.

The signal $x_{p_2,t}$ is defined as:

\begin{equation}\label{eq:OMDP2}
x_{p_2,t} (n) = s^{p_2}(n)\cdot w_t(n)
\end{equation}

where $s(n)$ is the speech signal and $w_t(n)$ is a Blackman window centered at time $t$:

\begin{equation}\label{eq:OMDP3}
w_t(n) = w(n-t)
\end{equation}

As in \cite{Drugman-GCI}, the window length is recommended to be proportionnal to the mean period $T_{0,mean}$ of the considered voice, so that $y_{p_1,p_2}(t)$ is almost a sinusoid oscillating at the local fundamental frequency. For $(p_1,p_2)=(1,1)$, the oscillating moment is the mean-based signal used in \cite{Drugman-GCI} for which the window length is $1.75 \cdot T_{0,mean}$. For oscillating moments of higher orders, we observed that a larger window is required for a better resolution. In the rest of this paper, we used a window length of $2.5 \cdot T_{0,mean}$ for higher orders (which in our analysis did not exceed 4). Besides, to avoid a low-frequency drift in $y_{p_1,p_2}(t)$, this signal is high-passed with a cut-off frequency of 40 Hz.

\begin{figure*}[!ht]
  \centering
  \includegraphics[width=1\textwidth]{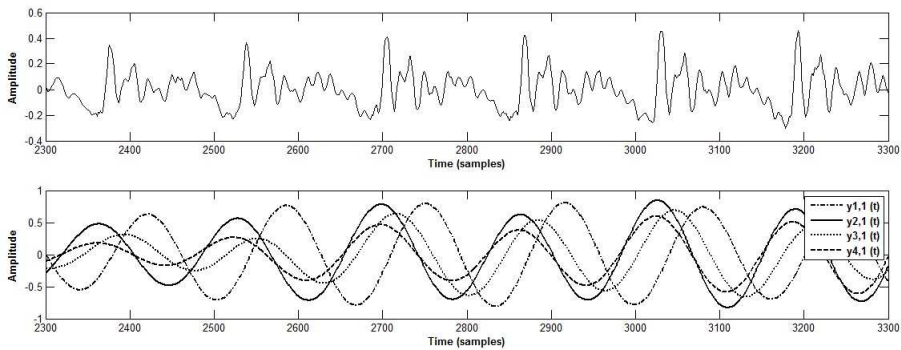}
  \caption{Illustration of the oscillating moments. \emph{Top plot}: the speech signal. \emph{Bottom plot}: the resulting oscillating moments with various values of $p_1$ and for $p_2 = 1$.}
  \label{fig:4oscillations}
\end{figure*}

Figure \ref{fig:4oscillations} illustrates for a given segment of voiced speech the evolution of four oscillating moments $y_{p_1,p_2}(t)$ respectively for $(p_1,p_2)=\{(1,1);(2,1);(3,1);(4,1)\}$. It can be noticed that all oscillating moments are sinusoids evolving at the local fundamental frequency and whose relative phase shift depends upon the order $p_1$. Note that a similar conclusion can be drawn when inspecting the effect of $p_2$. The principle of the proposed method is that $y_{p_1,p_2}(t)$ is polarity-dependent if $p_1 \cdot p_2$ is odd (i.e the oscillating moment is inverted with a polarity change), and is polarity-independent if $p_1 \cdot p_2$ is even. 

\begin{figure*}[!ht]
  \centering
  \includegraphics[width=1\textwidth]{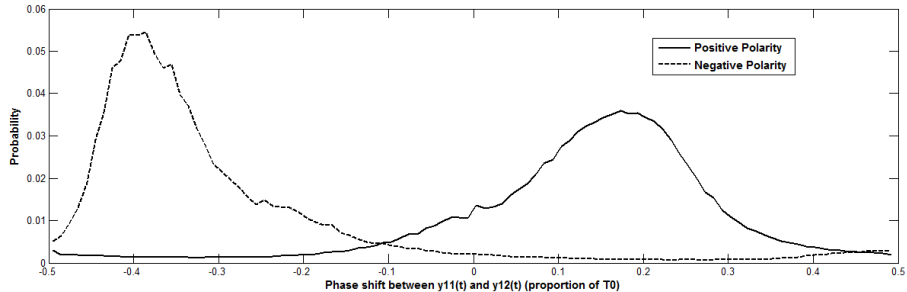}
  \caption{Distribution of the phase shift (in local pitch period) between $y_{1,1}(t)$ and $y_{1,2}(t)$ for a negative and positive polarity.}
  \label{fig:PhaseShift}
\end{figure*}

In the following tests, for the sake of simplicity, only the oscillating moments $y_{1,1}(t)$ and $y_{1,2}(t)$ (or $y_{2,1}(t)$) are considered. Figure \ref{fig:PhaseShift} shows, for the several speakers that will be analyzed in Section \ref{sec:Experiments}, how the distribution of the phase shift between $y_{1,1}(t)$ and $y_{1,2}(t)$ is affected by an inversion of polarity. Note that these histograms were obtained at the frame level and that phase shifts are expressed as a function of the local $T_0$. Figure \ref{fig:PhaseShift} suggests that fixing a threshold around -0.12 could lead to an efficient determination of the speech polarity.

Our proposed method, called Oscillating Moment-based Polarity Detection (OMPD), works as follows:

\begin{itemize}

\item Estimate roughly with an appropriate technique the \emph{mean} pitch value $T_{0,mean}$ (required for determining the window length) and the voicing boundaries.

\item Compute from the speech signal $s(n)$ the oscillating moments $y_{1,1}(t)$ and $y_{1,2}(t)$, as indicated by Equations \ref{eq:OMDP1} to \ref{eq:OMDP3}.

\item For each voiced frame, estimate the local pitch period $T_0$ from $y_{1,1}(t)$ (or equivalently from $y_{1,2}(t)$) and compute the local phase shift between these two signals.

\item Apply a majority decision over the voiced frames, a frame being with a positive polarity if its phase shift is comprised between -0.12 and 0.38. 

\end{itemize}

It is worth to mention that an important advantage of OMPD, with regard to the techniques described in Section \ref{sec:ExMethods}, is that it just requires a rough estimate of the mean pitch period, and not an accurate determination of the complete pitch contour. This gives the method also an advantage for performing in adverse conditions.

\section{Experiments}\label{sec:Experiments}
The experimental evaluation is carried out on 10 speech corpora. Several voices are taken from the CMU ARCTIC database \cite{CMUARCTIC}, which was designed for speech synthesis purpose: AWB (Scottish male), BDL (US male), CLB (US female), JMK (Canadian male), KSP (Indian male), RMS (US male) and SLT (US female). About 50 min of speech is available for each of these speakers. The Berlin database \cite{Berlin} is made of emotional speech (7 emotions) from 10 speakers (5F - 5M) and consists of 535 sentences altogether. The two speakers RL (Scottish male) and SB (Scottish female) from the CSTR database \cite{CSTR}, with around 5 minutes per speaker, are also used for the evaluation.

Results of polarity detection using the four techniques described in the previous sections are reported in Table \ref{tab:Results}. It can be noticed that GSGW gives in general a lower performance, except for speaker SB where it outperforms other approaches. PC generally achieves high detection rates, except for speakers SB and SLT. Although RPS leads to a perfect polarity determination in 7 out of the 10 corpora, it may for some voices (KSP and SB) be clearly outperformed by other techniques. As for the proposed OMPD method, it works perfectly for 8 of the 10 databases and gives an acceptable performance for the two remaining datasets. In average, over the 10 speech corpora, it turns out that OMPD clearly carries out the best results with a total error rate of 0.15\%, against 0.64\% for PC, 0.98\% for RPS and 3.59\% for GSGW.

\begin{table}[!ht]
\centering
\begin{tabular}{ c || c | c | c || c | c | c || c | c | c || c | c | c |}
    & \multicolumn{3}{|c||}{GSGW} & \multicolumn{3}{|c|}{PC} & \multicolumn{3}{|c|}{RPS} & \multicolumn{3}{|c|}{OMPD}\\       
   \hline
    Speaker & OK & KO & Acc. (\%) & OK & KO & Acc. (\%) & OK & KO & Acc. (\%) & OK & KO & Acc. (\%)\\  
    \hline
    \hline
    AWB & 1134 & 4 & 99.64 & 1138 & 0 & \textbf{100} & 1138 & 0 & \textbf{100} & 1138 & 0 & \textbf{100}\\  
    \hline
    BDL & 1112 & 19 & 98.32 & 1131 & 0 & \textbf{100} & 1131 & 0 & \textbf{100} & 1131 & 0 & \textbf{100}\\
    \hline
    Berlin & 356 & 179 & 66.54 & 528 & 7 & 98.69 & 535 & 0 & \textbf{100} & 525 & 10 & 98.13\\
    \hline
    CLB & 1131 & 1 & 99.91 & 1132 & 0 & \textbf{100} & 1132 & 0 & \textbf{100} & 1132 & 0 & \textbf{100}\\
    \hline
    JMK & 1096 & 18 & 98.38 & 1109 & 5 & 99.55 & 1114 & 0 & \textbf{100} & 1114 & 0 & \textbf{100}\\
    \hline
    KSP & 1103 & 29 & 97.43 & 1132 & 0 & \textbf{100} & 1059 & 73 & 93.55 & 1132 & 0 & \textbf{100}\\
    \hline
    RL & 50 & 0 & \textbf{100} & 50 & 0 & \textbf{100} & 50 & 0 & \textbf{100} & 50 & 0 & \textbf{100}\\
    \hline
    RMS & 1082 & 50 & 95.58 & 1132 & 0 & \textbf{100} & 1129 & 3 & 99.73 & 1132 & 0 & \textbf{100}\\
    \hline
    SB & 49 & 1 & \textbf{98} & 37 & 13 & 74 & 42 & 8 & 84 & 47 & 3 & 94\\
    \hline
    SLT & 1125 & 6 & 99.38 & 1101 & 30 & 97.35 & 1131 & 0 & \textbf{100} & 1131 & 0 & \textbf{100}\\
        \hline
        \hline
    \textbf{TOTAL} & 8238 & 307 & 96.41 & 8490 & 55 & 99.36 & 8461 & 84 & 99.02 & 8532 & 13 & \textbf{99.85}\\
    \hline
\end{tabular}
\newline
\caption{Results of polarity detection for 10 speech corpora using the four techniques. The number of sentences whose polarity is correctly (OK) or incorrectly (KO) determined are indicated, as well as the detection accuracy (in \%).}
\label{tab:Results}
\end{table}

%

\section{Conclusion}\label{sec:conclu}
This paper investigated the use of oscillating moments for the automatic detection of speech polarity. The proposed technique is based on the observation that these moments oscillate at the local fundamental frequency and have a phase shift which is dependent upon the speech polarity. The introduction of this polarity-dependency is made by considering non-linearity or higher-order statistics. The resulting method is shown through an objective evaluation on several large corpora to outperform existing approaches for polarity detection. On these databases, it reaches an average error rate of 0.15\% against 0.64\% for the best state-of-the-art technique. Besides the proposed method only requires a rough estimate of the \emph{mean} pitch period for the considered voice.

\section*{Acknowledgments}\label{sec:Acknowledgments}

Thomas Drugman is supported by the ``Fonds National de la Recherche Scientifique'' (FNRS).

\end{document}